\documentclass[%
 aps,superscriptaddress,
 amsmath,amssymb,
 reprint,%
]{revtex4-1}

\usepackage{graphicx}
\usepackage{dcolumn}
\usepackage{bm}
\usepackage{color,xcolor}
\usepackage[utf8]{inputenc}
\usepackage[T1]{fontenc}
\usepackage{mathptmx}
\usepackage{braket}
\usepackage[section]{placeins}

\newcommand{\be}{\begin{equation}}
\newcommand{\ee}{\end{equation}}

\newcommand{\bs}{\boldsymbol}

\newcommand{\revision}{}

\newcommand{\addcqt}{Centre for Quantum Technologies, National University of Singapore, 3 Science Drive 2, Singapore 117543}
\newcommand{\addtuc}{School of Electrical and Computer Engineering, Technical University of Crete, Chania, Greece 73100}

\begin{document}

\title{Photonic band structure design using persistent homology}

\author{Daniel Leykam}
\affiliation{\addcqt}
\author{Dimitris G. Angelakis}
\affiliation{\addcqt}
\affiliation{\addtuc}

\date{\today}

\begin{abstract}
The machine learning technique of persistent homology classifies complex systems or datasets by computing their topological features over a range of characteristic scales. There is growing interest in applying persistent homology to characterize physical systems such as spin models and multiqubit entangled states. Here we propose persistent homology as a tool for characterizing and optimizing band structures of periodic photonic media. Using the honeycomb photonic lattice Haldane model as an example, we show how persistent homology is able to reliably classify a variety of band structures falling outside the usual paradigms of topological band theory, including ``moat band'' and multi-valley dispersion relations, and thereby control the properties of quantum emitters embedded in the lattice. The method is promising for the automated design of more complex systems such as photonic crystals and Moir\'{e} superlattices. 
\end{abstract}

\maketitle

\section{Introduction}

There is growing interest in the design of synthetic gauge fields and topological band structures for light~\cite{Haldane2008,Wang2009,Malkova2009,Khanikaev2013,Rechtsman2013,Hafezi2013}, motivated both by their exotic behaviour and potential applications as disorder-robust photonic devices~\cite{topo_review,topo_review2}. The most common approach has been to emulate well-known topological phases from condensed matter physics~\cite{TI_review}. However, recent advances in synthetic dimensions in photonic systems provide platforms for observing for the first time topological effects in higher dimensions~\cite{synthetic1,synthetic2,synthetic3}. Given the high degree of control and flexibility we now have over the design of photonic band structures and synthetic gauge fields, with many different degrees of freedom, a natural question is whether there exists interesting and useful topological phenomena falling outside the well-established paradigm of topological band theory, which is concerned with the ``shape'' of Bloch wave eigenstates, i.e. whether they exhibit vortices or gauge discontinuities within the Brillouin zone, independent of the energy dispersion.

Controlling the shape of the energy dispersion landscape can be equally important. For example, the shape of a medium's isofrequency contours or surfaces dictates the far-field radiation profile of localized emitters~\cite{hyperbolic_review}, and flat dispersion relations are highly desirable for realizing strong light-matter interactions~\cite{FB_review}. Both of these examples are based on the band structure at a specific energy. However, perfectly flat Bloch bands are idealizations never achievable in practice, and real systems inevitably exhibit losses, meaning there will be a finite energy resolution. These considerations motivate the use of ``fuzzy'' tools for characterizing the topology and abstract ``shapes'' of photonic media.

The characterization of topological properties in the presence of finite resolution or noise can be carried out using persistent homology, which is a tool from the field of topological data analysis (TDA)~\cite{Ghrist2008,TDA_review,TDA_review2}. Persistent homology characterizes the ``shape'' and connectivity of high-dimensional data sets by studying how their topological properties change as a characteristic distance scale is varied. Several studies have started to apply persistent homology and other techniques from TDA to physics problems~\cite{Murugan2019,Mengoni2019,Spitz2020,Tran2020,Olsthoorn2020,Cole_arxiv}. For example, persistent homology has recently been used to characterize entanglement of multiqubit states~\cite{Mengoni2019}, identify different dynamical regimes of the Bose-Hubbard model~\cite{Spitz2020}, and recognize topological phase transitions in models of interacting spins~\cite{Tran2020,Olsthoorn2020,Cole_arxiv}. To apply persistent homology, the important first step is to choose a suitable distance metric to characterize the data.

Our aim in this study is introduce persistent homology to photonics and show how it may be useful for optimization of synthetic gauge fields and discovery of novel classes of photonic band structures which merit more detailed studies. As a simple example, we consider the characterization of the Haldane model, a tight binding model describing a honeycomb array of waveguides or resonators in the presence of a synthetic gauge field~\cite{Haldane0,Haldane1,Haldane2,Haldane3}. We show how to use persistent homology to characterize the shape of its isoenergy contours and the connectivity of the Bloch function eigenstates in the vicinity of this energy, described by the quantum distance~\cite{Haldane_talk,Kolodrubetz2013,Palumbo2018,Bleu2018,Rhim2020,Gianfrate2020,SCQ}. As an application of this approach, we will show how these two features can be used to optimize the lifetimes and radiation profiles of embedded quantum emitters. 

The outline of this article is as follows: In Sec.~\ref{sec:methods} we provide a brief introduction to the technique of persistent homology and discuss how it may be applied to band structure optimization by using suitable distance metrics. Sec.~\ref{sec:haldane} applies persistent homology to analyze the band structure of the Haldane model, identifying parameter ranges exhibiting ``moat band''~\cite{Sedrakyan2014,Sedrakyan2015} and multivalley dispersion relations~\cite{multivalley,multivalley2}. In Sec.~\ref{sec:emitter} we show how the considered distance metrics can be used to predict properties of quantum emitters embedded in the lattice. Sec.~\ref{sec:conclusion} concludes with discussion of possible future directions and applications of persistent homology in photonics. 

\section{Methods}
\label{sec:methods}

\subsection{A brief introduction to persistent homology}

The following discussion is quite condensed and not intended to be a comprehensive introduction to persistent homology. For a more pedagogical and complete introduction we recommend the excellent Ref.~\cite{Murugan2019} aimed at physicists, or other recent articles applying persistent homology to characterize quantum and condensed matter systems~\cite{Mengoni2019,Spitz2020,Tran2020,Olsthoorn2020}.

\begin{figure}
    \centering
    \includegraphics[width=0.8\columnwidth]{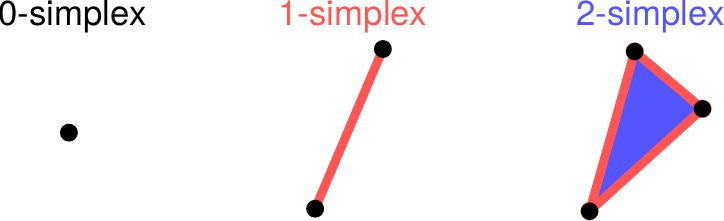}
    \caption{Examples of $k$-simplices. The edges of a $k$ simplex are $(k-1)$-simplices.}
    \label{fig:simplices}
\end{figure}

Persistent homology is a technique used to characterize sets of data $\{ \bs{x}_1, \bs{x}_2,...,\bs{x}_N \}$ equipped with some measure of distance between them $d(\bs{x}_i,\bs{x}_j) \geq 0$. For example, the data $\bs{x}_i$ could correspond to particle positions in real space characterized by the Euclidean metric $d(\bs{x}_i,\bs{x_j}) = |\bs{x}_i - \bs{x}_j|$, or image data with $\bs{x}_i$ encoding the locations and intensities of the individual pixels. Given the data, one can construct subsets known as $k$-simplices, where each $k$-simplex is a set of $k+1$ of the points; 0-simplices are individual points $\{\bs{x}_i\}$, 1-simplices are edges connecting pairs of points $\{\bs{x}_i,\bs{x}_j\}$, 2-simplices are areas enclosed by three points $\{\bs{x}_i,\bs{x}_j,\bs{x}_k\}$, and so on, as illustrated in Fig.~\ref{fig:simplices}.

To characterize the shape of the data, we construct sets of simplices using the distance measure $d$ to form \emph{simplicial complexes}. A $k$-complex is a closed set of simplices of dimension up to $k$. By closed, we mean that if a simplex belongs to the complex, then its edges are also in the complex. For example, if an edge $\{\bs{x}_i,\bs{x}_j\}$ belongs to the complex, then its constituent points $
\{\bs{x}_i\}, \{\bs{x}_j\}$ also belong to the complex. In the case of Euclidean data a common choice is the Vietoris-Rips complex $\mathcal{V}_{\epsilon}$, which includes all $k$-simplices whose constituent points all have a pairwise distance less than $\epsilon$ from each other.

By counting the number of simplices of each dimension in the complex we can determine its abstract ``shape''. For example, a complex consisting of 3 points and 2 edges describes a line, while a complex with 3 points and 3 edges forms a loop. Formally, the existence of topologically nontrivial structures of dimension $k$ in the simplicial complex is determined by computing its $k$th Betti number $B_k$, which corresponds to the number of $k$-dimensional ``holes'' in the complex.

Adding a single vertex or edge to the complex can drastically alter its topological properties encoded by the Betti numbers. They key innovation provided by persistent homology which makes it useful for characterizing real world systems with noise is to study how the topology of the simplicial complex varies with some characteristic scale $\epsilon$, in effect studying the shape of the data over a range of scales, known as a \emph{filtration}. Features which only persist for a small range of scales can be attributed to noise and discarded, while those that persist for a large range of scales are robust and meaningful.

There are several possible choice for the filtration, depending on the data to be analyzed. For the case of point cloud data, one can use the Vietoris-Rips complex $\mathcal{V}_{\epsilon}$. However, this becomes extremely time consuming to compute for large numbers of data points. For point cloud data there are other, more efficient complexes such as the $\alpha$ complex~\cite{Ghrist2008,TDA_review}. For image data (pixels on a regular grid), another choice is the sublevel set filtration constructed out of all neighbouring pixels with intensities less than (or greater than) some threshold value $\epsilon$, which characterizes the number and shape of intensity minima (or maxima). 

Given a dataset, distance metric, and choice of filtration, the persistent homology of the data can be computed using existing software libraries~\cite{Julia,Eirene,Ripserer}. As long as the dataset is not too large, low dimensional topological features can be computed relatively quickly. However, computation time grows rapidly for high-dimensional features, generally requiring pre-processing to reduce the dataset size. 

\subsection{Characterization of energy bands using persistent homology}

Let us now return to physics. In order to apply persistent homology to physical systems, we first need to identify the relevant data set $\{ \bs{x}_i \}$ and a suitable distance metric to characterize the data. Several different approaches have recently been proposed in the physics literature~\cite{Murugan2019,Mengoni2019,Spitz2020,Tran2020,Olsthoorn2020}. Here our aim is to use to persistent homology to characterize energy bands of periodic photonic media. 

We will consider for simplicity Bloch wave spectra of photonic waveguide lattices obtained from the Schr\"odinger eigenvalue problem~\cite{topo_review}
\be 
\hat{H}(\bs{k}) \ket{u_n(\bs{k})} = \omega_n(\bs{k})\ket{u_n(\bs{k})}, \label{eq:SE}
\ee
where $\hat{H}(\bs{k})$ is the lattice's Bloch Hamiltonian, $\bs{k}$ is the Bloch momentum (restricted to the first Brillouin zone), $\ket{u_n(\bs{k})}$ is the Bloch function, with the bra-ket notation encoding the internal (e.g. sublattice or spin) degrees of freedom, $n$ is the band index, and $\omega_n$ is the frequency of the mode. We note that Maxwell's equations for electromagnetic modes of periodic photonic media can be cast into a similar form to Eq.~\eqref{eq:SE}, making our approach equally applicable to the design and optimization of photonic crystals. 

Numerical solutions of Eq.~\eqref{eq:SE} typically sample $\bs{k}$ on a regular grid in the Brillouin zone, providing a discrete set of data (the eigenvectors and eigenvalues) to which we can apply persistent homology. Keeping with the spirit of successful applications of machine learning to physics, our aim is to employ simple distance metrics with well-established importance that can be easily interpreted.

First, we consider classifying the energy eigenvalues of a band $\omega(\bs{k})$ using a similar approach to that used for characterizing image data. That is, we sample $\omega$ \revision{on a $\bs{k}$ mesh spanning the Brillouin zone. This mesh can be a regular grid respecting the lattice symmetries (e.g. triangular, cubic, etc.), or even a random grid, provided a sufficiently large number of points are sampled.} We then define a filter function $f(\bs{k}) = |\omega(\bs{k}) - \omega_0|$, where $\omega_0$ is some reference frequency of interest. Persistent homology can characterise the shape of the set of momenta $\bs{k}$ whose Bloch mode frequencies are within $\epsilon_{\omega}$ of the reference frequency $\omega_0$, i.e.
\be 
\kappa = \{ \bs{k} \; : \; |\omega(\bs{k}) - \omega_0 | < \epsilon_{\omega} \}.
\ee
In particular, persistent homology can tell us the range of frequency resolutions $\epsilon_{\omega}$ over which topological features of the lattice's isoenergy contours persist.

An important consideration is that we inevitably sample $\bs{k}$ on a discrete grid of length $N$; hence there will be noise induced by the discretization. To be precise, neighbouring grid points will have an energy difference $\delta \omega \approx \bs{v}_g \cdot \delta \bs{k}$, where $\bs{v}_g = \nabla_{\bs{k}} \omega$ is the local wave group velocity, and $\delta k = 2\pi / N$ is the grid spacing (using units such that the lattice period $a=1$). Thus, topological features of the isoenergy surfaces are only meaningful if they persist over a range of energies larger than $\delta \omega = \mathrm{max}_{\kappa} v_g / \delta k$. As a crude estimate, if $\Delta$ is the width of the band of interest, the group velocity will have a scale $v_g \sim \frac{\Delta }{2\pi}$. Therefore as a rule of thumb one should look for features with a minimum persistence of $\Delta / N$ to avoid discretization errors; in the following we will use $\delta \omega = 2\mathrm{max}_{\kappa}v_g /N$ as the persistence cutoff.

The calculation proceeds as follows: (1) Choose a reference energy $\omega_0$ and grid size $N$. (2) Compute the persistent homology of the isoenergy contours, using the energy resolution $\epsilon_{\omega}$ as the filter function. (3) Discard features with persistence less than $\delta \omega$. (4) Sum the number of remaining features existing at each $\epsilon_{\omega}$ of interest to obtain the Betti numbers, which characterize the shape of the isoenergy surface at that  resolution, i.e. its connectivity and number of holes.

\begin{figure}
    \centering
    \includegraphics[width=\columnwidth]{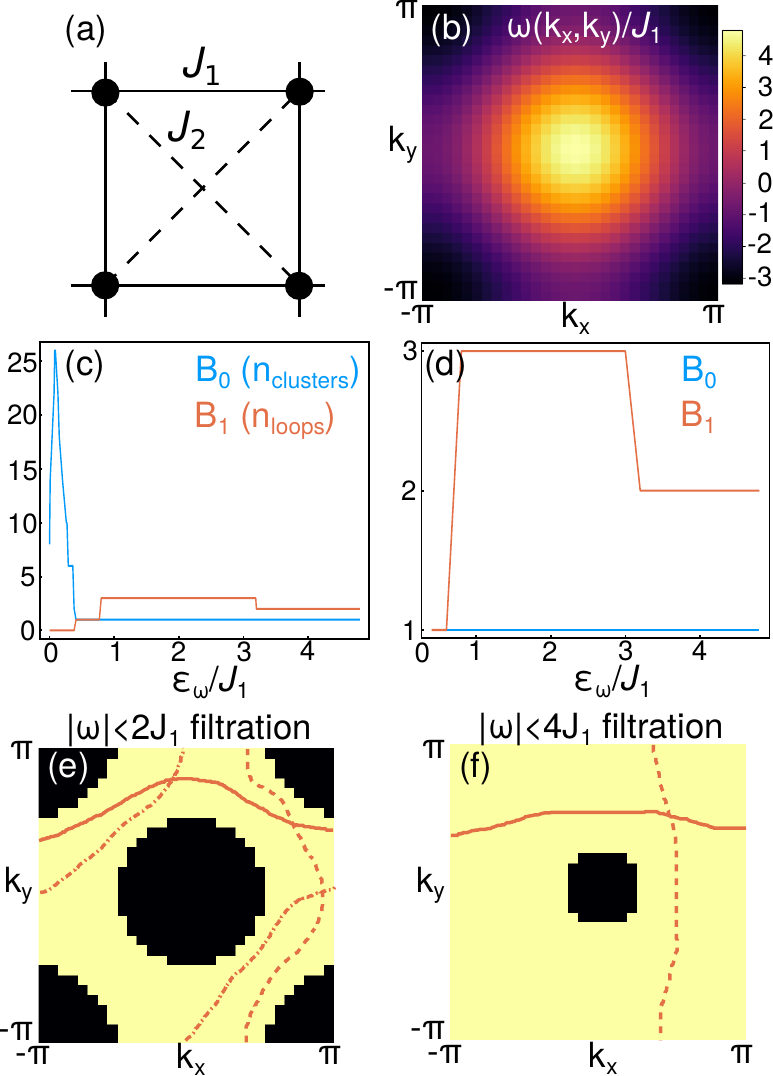}
    \caption{Persistent homology of an energy band. (a) Simple square lattice with nearest neighbour coupling $J_1$ and next-nearest neighbour coupling $J_2$. (b) Energy dispersion $\omega(\bs{k})$ for $J_2 = 0.2J_1$ sampled on an $N\times N = 30\times 30$ Brillouin zone grid. (c) Betti numbers $B_{0,1}$ obtained from the filtration $|\omega(\bs{k})|<\epsilon_{\omega}$. (d) Betti numbers obtained when retaining only features with a persistence greater than the discretization-induced noise. (e,f) Representative \revision{sublevel set} filtrations in the two phases. Momenta used to construct the simplicial complex are shaded in beige. Representative cycles are denoted by the orange lines. (e) $|\omega(\bs{k})|<2J_1$, $B_1 = 3$. (f) $|\omega(\bs{k})|<4J_1$, $B_1 = 2$.}
    \label{fig:energy_example}
\end{figure}

As an example of this procedure, we consider the energy band of a simple square lattice shown in Fig.~\ref{fig:energy_example}(a), described by the dispersion relation
\be 
\omega(\bs{k}) = 2J_1 (\cos k_x + \cos k_y) + 2J_2 (\cos [k_x+k_y] + \cos [k_x-k_y]),
\ee
where $J_1$ and $J_2$ are the nearest and next-nearest neighbour hopping strengths, respectively. We take the reference energy to be $\omega_0 = 0$, close to the band centre, and $J_2 = 0.2J_1$. The dispersion $\omega(\bs{k})$ is shown in Fig.~\ref{fig:energy_example}(b).

Applying persistent homology to this two-dimensional dispersion ``image'', we obtain the energy resolution-dependent Betti numbers $B_{0,1}(\epsilon_{\omega})$ shown in Fig.~\ref{fig:energy_example}(c). For small $\epsilon$ there is a sharp peak in the number of distinct clusters $B_0$, which is induced by discreteness of the momentum grid and highly sensitive to the chosen grid size $N$. From this plot we see that thresholding, i.e. consideration of features only with a minimum persistence, is essential to obtaining meaningful Betti numbers.

Applying the minimum persistence threshold in Fig.~\ref{fig:energy_example}(d), we obtain two ``phases'' with different 1st Betti numbers $B_1$, which we counts the numbers of loops (or holes) in the data. Fig.~\ref{fig:energy_example}(e) illustrates the first phase, considering modes with energies $|\omega| < 2J_1$, where $2J_1$ is the chosen energy resolution. $B_0 = 1$ indicates the modes form a single connected cluster, as can be readily verified. $B_1 = 3$ means that the image hosts three inequivalent non-contractible loops, as shown. We note that one needs to be careful with counting non-contractible loops here, due to the periodic boundary conditions of the Brillouin zone. 

As the energy cutoff $\epsilon_{\omega}$ is increased, the band minimum at $\omega \approx -3J_1$ becomes included in the filtration, destroying one of the holes and resulting in the image shown in Fig.~\ref{fig:energy_example}(f). The loss of a hole corresponds to $B_1$ decreasing by 1. Note that that the removal of the remaining hole in the centre of the image does not affect $B_1$; due to the Brillouin zone's periodic boundary conditions, any loop encircling this hole can be continuously deformed to a point by first expanding it to the Brillouin zone edges at $k_x = \pm \pi$ to create a pair of loops, and then annihilating the pair at the $k_y = \pm \pi$ Brillouin zone edge.

\subsection{Characterization of Bloch functions}

Persistent homology allows us to also classify more complicated objects than energy band structure ``images''. In particular, topological phases of light and matter are associated with non-trivial topological properties of the Bloch function eigenstates $\ket{u_n(\bs{k})}$, which are usually computed using the Berry connection or curvature~\cite{topo_review}. Here we will consider the abstract shape of the Bloch functions characterized by the quantum distance $d$,
\be
d^2(\bs{k},\bs{k}^{\prime}) = 1 - |\braket{u_n(\bs{k})|u_n(\bs{k}^{\prime})}|^2.
\ee
The quantum distance between Bloch functions at infinitesimally-separated momenta corresponds to the quantum metric tensor $\hat{g}$, with components
\be 
g_{ij} = \mathrm{Re}\left[ \sum_{m \neq n} \frac{ \bra{u_m} \partial_{k_i} \hat{H}(\bs{k}) \ket{u_n} \bra{u_n} \partial_{k_j} \hat{H}(\bs{k}) \ket{u_m}}{(E_m - E_n)^2} \right].
\ee 
The study of interesting observables related to properties of the quantum distance and quantum metric currently attracts broad interest~\cite{Haldane_talk,Kolodrubetz2013,Palumbo2018,Bleu2018,Rhim2020,Gianfrate2020,SCQ}. For example, in graphene the large quantum distance between counter-propagating modes at the Dirac cones is responsible for the suppression of backscattering in the presence of long range (valley-conserving) disorder~\cite{Haldane_talk}; a small quantum distance means there can be efficient scattering between the two states by scalar (e.g. spin-preserving) disorder.

We use the quantum distance to define a graph, with each vertex corresponding to a Bloch function with some momentum $\bs{k}$. To apply persistent homology, we define two points $\bs{k}$, $\bs{k}^{\prime}$ to be connected when $d(\bs{k},\bs{k}^{\prime}) < \epsilon_d$. The Betti numbers of the resulting graph can then be used to characterize the ``shape'' of the eigenstates, such as the existence of loops or multiple disconnected clusters. One can either consider how the number of features changes as a function of $\epsilon_d$, or consider the features at a particular $\epsilon_d$.

As with the energy eigenvalues, sampling of the Brillouin zone on a discrete grid will induce discretization noise for small quantum distance cutoffs $\epsilon_d$. For neighbouring grid points, $d(\bs{k},\bs{k}+\delta \bs{k}) \approx \sqrt{\delta \bs{k} \cdot \hat{g}(\bs{k}) \cdot \delta \bs{k}}$. Therefore to avoid discretization errors one should consider features that persist for $\gtrsim \frac{2\pi}{N} \sqrt{\hat{g}}$. 

We note that under this definition, the existence of non-trivial shapes of the energy eigenvalues is a necessary (but not sufficient) condition for the Bloch functions to have non-trivial shapes, since the Bloch functions are typically continuous functions of $\bs{k}$. Furthermore, while the energy eigenvalues share the periodicity of the reciprocal lattice, i.e. $\omega_n(\bs{k}) = \omega_n(\bs{k}+\bs{G})$ for any reciprocal lattice vector $\bs{G}$, the quantum distance generally does \emph{not} share this periodicity. Thus, $\bs{d}(\bs{k},\bs{k}+\bs{G})$ can be nonzero. In the following we will compute topological features of the Bloch functions in the first Brillouin zone.

\begin{figure}
    \centering
    \includegraphics[width=\columnwidth]{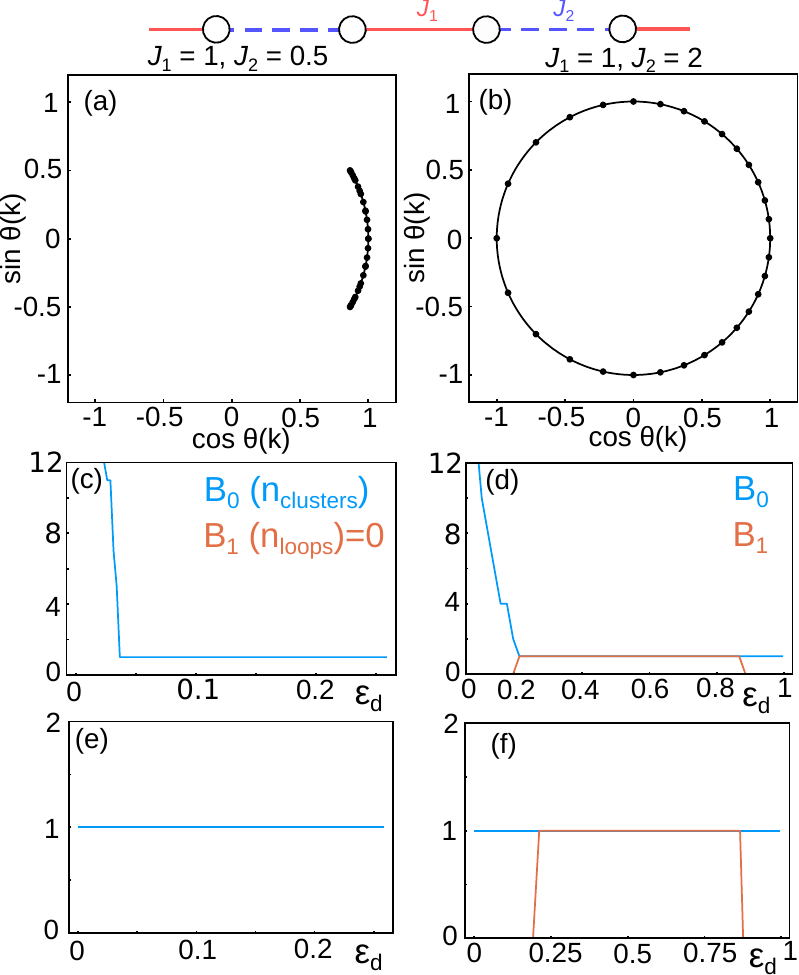}
    \caption{Persistent homology of Bloch functions in the Su-Schrieffer-Heeger model in the trivial (a,c,e) and nontrivial (b,d,f) phases. (a,b) Representation of Bloch functions as points on the unit circle. Solid lines denote the continuum limit, dots indicate the numerically-sampled points. (c,d) Betti numbers obtained using a filtration based on the quantum distance between Bloch functions at momenta $k$ and $k^{\prime}$, $d(k,k^{\prime})$. (e,f) Filtrations after discarding features with persistence less than the discretization-induced noise.}
    \label{fig:SSH}
\end{figure}

As an example, we consider the connectivity of the Bloch function eigenstates in the one-dimensional Su-Schrieffer-Heeger (SSH) model shown in Fig.~\ref{fig:SSH}, which is described by the Bloch Hamiltonian~\cite{topo_review}
\be 
\hat{H}(k) = (J_1 + J_2 \cos k)\hat{\sigma}_x + J_2 \sin k \hat{\sigma}_y, \label{eq:ssh}
\ee 
where $J_{1,2}$ are coupling strengths and $\hat{\sigma}_i$ are Pauli matrices. The eigenvectors of Eq.~\eqref{eq:ssh} are 
\begin{align}
&\ket{u_{\pm}(k)} = \frac{1}{\sqrt{2}} (1, \pm e^{-i \theta(k)})^T,\\
\theta(k) &= \mathrm{Arg}(J_1 + J_2 \cos k - i J_2 \sin k),
\end{align}
and lie on the unit circle. The eigenvectors span the entire circle in the nontrivial phase $J_2 > J_1$, and are limited to $-\pi/2 < \theta < \pi/2$ in the trivial phase. The quantum distance between eigenstates at $k$ and $k^{\prime}$ can be evaluated exactly as
\be 
d^2(k,k^{\prime}) =  \sin^2 \left( \frac{\theta(k)-\theta(k^{\prime})}{2} \right),
\ee
attaining its maximal value of 1 only in the nontrivial phase. We sample the Brillouin zone on a uniform grid $k = k_n = 2\pi n / N$, using $N=30$ grid points. Figs.~\ref{fig:SSH}(c,d) show the Betti numbers obtained from the two phases, without using any persistence threshold. Similar to the case of the energy eigenvalue images, discretization-induced noise appears for small filtration values $\epsilon_d$ as $N$-dependent peaks in $B_0$. 

To systematically remove this noise, we introduce a persistence threshold based on the quantum metric, which in the SSH model takes the form
\be 
g_{kk} = \lim_{dk\rightarrow0} \frac{d(k,k+dk)^2}{dk^2} = \frac{J_2^2 (J_2 + J_1 \cos k)^2}{4(J_1^2 + J_2^2 + 2 J_1 J_2 \cos k )^2}.
\ee
Using a threshold of $2.5\sum_n \frac{2\pi}{N^2} \sqrt{g_{kk}(k_n)}$ (i.e., proportional to the average quantum distance), we obtain smooth curves distinguishing the trivial and nontrivial phases in Fig.~\ref{fig:SSH}(e,f). In particular, the winding of the Bloch functions in the nontrivial phase is characterized by $B_1 = 1$, which persists over a large range of scales $\epsilon_d$. We note that as $\epsilon_d \rightarrow 1$ each point becomes connected to all others, destroying the hole and resulting in $B_1 = 0$. Thus, to characterize the shape of the Bloch functions one should use an intermediate value of $\epsilon_d$.

\section{Classifying band minima in the Haldane model}
\label{sec:haldane}

\begin{figure}
    \centering
    \includegraphics[width=\columnwidth]{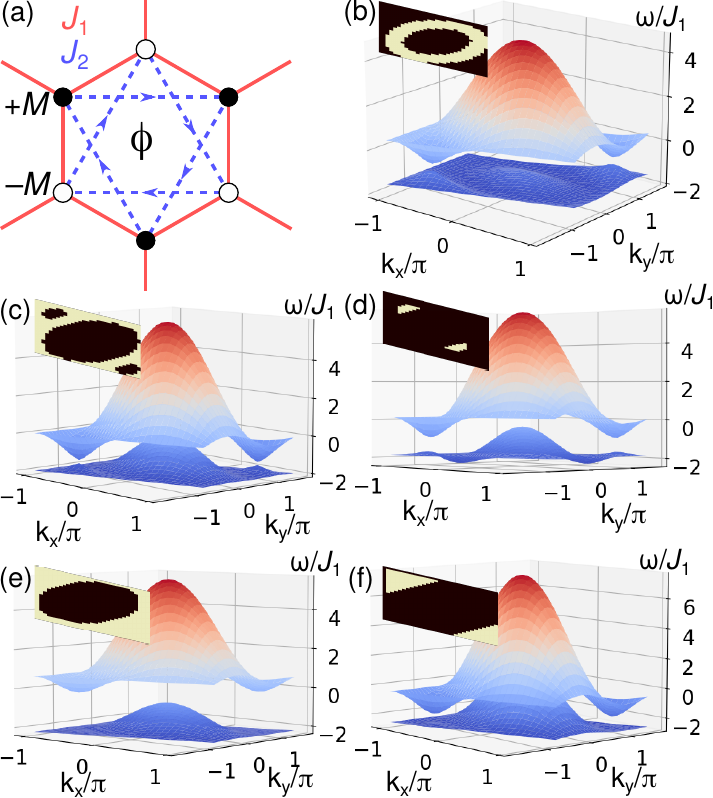}
    \caption{(a) Schematic of the Haldane model. (b-f) Band structures and \revision{sublevel set filtrations} (light regions in insets) for parameter values $(J_2,M,\phi)$ in different phases identified using persistent homology. (b) Single moat band ($0.3 J_1,0,0.3$). (c) Multiply-connected moat band with $B_1 = 4$ ($0.5 J_1,0,0.1$). (d) Two isolated valleys ($0.5 J_1,0,0.5$). (e) Moat band with $B_2 = 2$ ($0.4 J_1, J_1,0$). (f) Boundary between different phases ($0.75J_1,0.33J_1,0$).}
    \label{fig:haldane_schematic}
\end{figure}

We now apply the above measures to the classification of the Haldane model shown in Fig.~\ref{fig:haldane_schematic}(a)~\cite{Haldane0,Haldane1,Haldane2,Haldane3}, described by the Bloch wave Hamiltonian
\begin{align}
\hat{H}(\bs{k}) = & 2 J_2 \cos \phi \sum_{i=1}^3 \cos (\bs{k}\cdot \bs{a}_i) \hat{\sigma}_0 + J_1 \sum_{i=1}^3 \left[ \cos (\bs{k}\cdot \bs{\delta}_i) \hat{\sigma}_x \right. \label{eq:haldane} \\ 
& \left. + \sin(\bs{k}\cdot \bs{\delta}_i) \hat{\sigma}_y \right] + [M + 2 J_2 \sin \phi \sum_{i=1}^3 \sin (\bs{k}\cdot \bs{a}_i) ] \hat{\sigma}_z, \nonumber
\end{align}
where $\bs{\delta}_1 = (0,1/\sqrt{3})$, $\bs{\delta}_{2} = \frac{1}{2}(-1,-1/\sqrt{3})$, and $\bs{\delta}_3 = \frac{1}{2}(1,-1/\sqrt{3})$ are displacements between neighbouring unit cells, $\bs{a}_1 = (1,0)$, and $\bs{a}_2 = \frac{1}{2}(1,\sqrt{3})$, and $\bs{a}_3 = \frac{1}{2}(-1,\sqrt{3})$ are the lattice vectors, $J_{1}, J_{2} \geq 0$ are nearest- and next-nearest-neighbour hopping strengths, $M$ is the sublattice detuning, and $\phi$ parameterizes the staggered next-nearest-neighbour hopping phases.

A previous study showed that when $M = \phi = 0$, the Haldane model can exhibit ``moat bands'', i.e. degenerate band minima forming closed lines in the Brillouin zone~\cite{Sedrakyan2014}. The existence condition for moat bands when $M = \phi = 0$ was obtained by rewriting $\hat{H}(\bs{k})$ as
\begin{align}
& \hat{H}(\bs{k}) = J_1 \hat{T} + J_2 \hat{T}^2,\\ 
\hat{T} = T_x \hat{\sigma}_x + T_y \hat{\sigma}_y &= \sum_i \left[ \cos (\bs{k}\cdot \bs{\delta}_i) \hat{\sigma}_x + \sin(\bs{k}\cdot \bs{\delta}_i) \hat{\sigma}_y \right], \nonumber
\end{align}
which implies its energy eigenvalues take the form $\omega_{1,2} = \pm J_1 \sqrt{ T_x^2 + T_y^2} + J_2 (T_x^2 + T_y^2)$. It follows that the eigenvalues exhibit a degenerate minimum along the contour $\sqrt{T_x^2+T_y^2} = J_1/(2J_2)$. Since $\sqrt{T_x^2+T_y^2} \leq 3$, moat bands can only exist when the next-nearest neighbour coupling strength is sufficiently strong, i.e. $J_2 > J_1/6$. When a moat band is formed the ground state of hard-core bosons at low densities spontaneously breaks time-reversal symmetry~\cite{Sedrakyan2014}.

\begin{figure}
    \centering
    \includegraphics[width=\columnwidth]{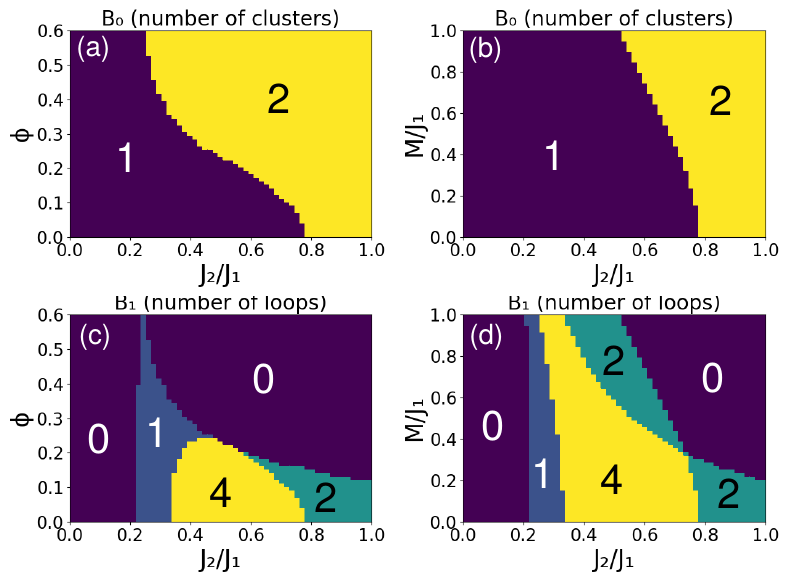}
 
    \caption{Betti numbers \revision{$B_0$ (a,b) and $B_1$ (c,d)} of the energy band minima of the Haldane model as a function of $J_2$ and $\phi$ (a,c) and $J_2$ and $M$ (b,d), obtained by applying persistent homology to the energy eigenvalues with a resolution $\epsilon_{\omega} = 0.1 J_1$. Persistent homology captures the emergence and changes in the moat band minimum known for $\phi=0$, and demonstrates its persistence for nonzero $\phi$. For large $\phi$ or $M$ and $J_2/J_1$ the moat is destroyed and replaced with a pair of isolated minima.}
    \label{fig:haldane_figure_1}
\end{figure}

When $M, \phi \neq 0$ the simple condition for the existence of a moat band no longer holds. We apply the persistent homology approach discussed in \revision{Sec. IIB} to systematically determine the fate of the moat band for $M,\phi \neq 0$. To do so, we set the reference energy as $\omega_0 = \mathrm{min}_{\bs{k}} \; [\omega_1(\bs{k})]$, considering the shape of the lowest energy eigenvalues of $\hat{H}(\bs{k})$. We sample the Brillouin zone on a grid of $N\times N = 30 \times 30$ points and consider features at a fixed frequency resolution $\epsilon_{\omega} = 0.1 J_1$. Figs.~\ref{fig:haldane_schematic}(b-f) show examples of the band structure in different phases identified using persistent homology, exhibiting different shapes of the lowest energy modes characterized by the Betti numbers plotted in Fig.~\ref{fig:haldane_figure_1}.

First, when $M=\phi = 0$ we observe that the persistent homology approach accurately reproduces the features of the energy minima obtained in Ref.~\cite{Sedrakyan2014}. Note that the threshold for the emergence of a single loop is slightly larger than the analytical threshold $J_1 = J_1/6$, due to the finite frequency resolution we consider, which results in a filled disk for small loop radii. Fig.~\ref{fig:haldane_schematic}(b) shows an example of a single-looped energy minimum ($B_1 = 1$), which persists for nonzero $\phi$ or $M$. 
As $J_2$ is increased the loop expands, generating additional loops once it intersects the edges of the Brillouin zone, characterized by Betti numbers $B_1 = 4$ [Fig.~\ref{fig:haldane_schematic}(c)] or $2$ [Fig.~\ref{fig:haldane_schematic}(e)], depending on the parameter values. For large $J_2$ the single connected contour splits into two isolated ring-shaped energy minima centred at the two $\bs{K}$ points. The differing behaviour under increasing $\phi$ and $M$ is noteworthy; increasing $\phi$ always results in the destruction of the loops and their replacement with a pair of isolated, point-like energy minima [Fig.~\ref{fig:haldane_schematic}(d)], whereas the loops can persist in the presence of moderate $M$.

Interestingly, by fine-tuning a pair of parameters $(J_2,\phi)$ or $(J_2,M)$, one can even obtain a ``phase transition point'' at which multiple Betti numbers $B_{0,1}$ intersect [Fig.~\ref{fig:haldane_schematic}(f)]. At this point it is possible to obtain a variety of shapes of the low energy modes by applying small perturbations.

It is also noteworthy that for the chosen energy resolution $\epsilon_{\omega} = 0.1J_1$, the nontrivial low energy mode shapes emerge when the lower band is much flatter than the upper band, making the shapes difficult to directly identify from the images of the full band structure in Fig.~\ref{fig:haldane_schematic}. One could alternatively normalize the energy resolution by the width of the band of interest, or consider features at other energies. For example, if $\omega_0$ is set to the band gap, the nonzero overlap between the upper and lower bands in Figs.~\ref{fig:haldane_schematic}(c,d,f) can generate multiple disconnected loops with $B_{0,1} > 1$. 

\begin{figure}
    \centering
    \includegraphics[width=\columnwidth]{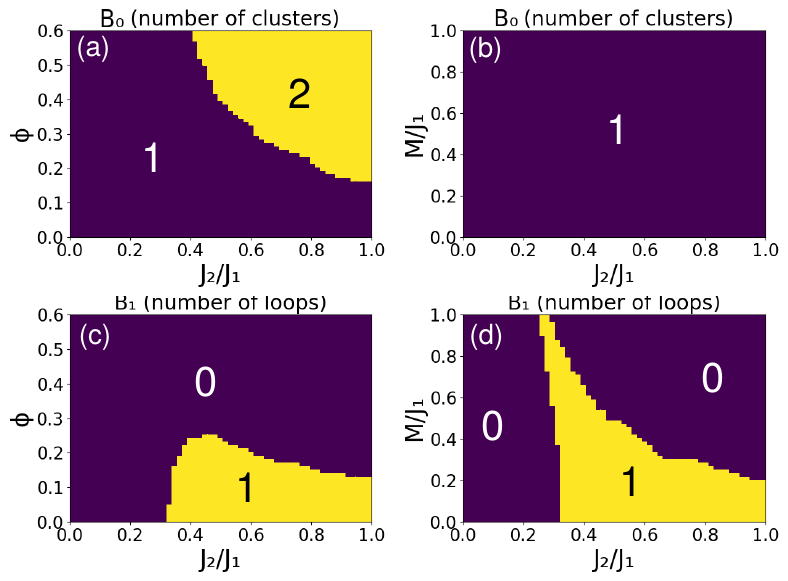}
 
    \caption{Betti numbers \revision{$B_0$ (a,b) and $B_1$ (c,d)} of the low energy eigenstates of the Haldane model using a resolution $\epsilon_d = 0.5$ \revision{as a function of $J_2$ and $\phi$ (a,c) and $J_2$ and $M$ (b,d)}. For large $\phi$ and $J_2$ the moat is destroyed and replaced with a pair of isolated minima with degenerate energies but near-orthogonal Bloch functions, forming two distinct valleys. On the other hand, for large $M$ and $J_2$ the two minima exhibit similar Bloch functions, separated by a small quantum distance.}
    \label{fig:haldane_figure_2}
\end{figure}

Next, in Fig.~\ref{fig:haldane_figure_2} we study the connectivity of the low energy Bloch functions characterized by the quantum distance. We consider features at $\epsilon_d = 0.5$, motivated by the SSH model example discussed in the previous Section. \revision{The Bloch function connectivity provides additional information about the shape of the band properties, complementary to the shape of the low energy eigenvalues in Fig.~\ref{fig:haldane_figure_1}. Since the Bloch functions are continuous functions of the momentum $\bs{k}$, they can only form disconnected clusters if the isoenergy contours also form disconnected clusters. This explains why the $B_0=2$ states appearing in Fig.~\ref{fig:haldane_figure_2}(a) are fully contained in the $B_0 = 2$ region of Fig.~\ref{fig:haldane_figure_1}(a). Similarly for the case of $B_1$, isoenergy loops are neccessary for loops of the Bloch functions to appear.}

Moderate $\phi$ and $M$ have qualitatively different effects on the shape of the Bloch functions; moderate $\phi$ results in the creation of two disconnected clusters of eigenvectors, while under increasing $M$ the Bloch functions only form a single cluster. To understand these differing behaviours, we note that at large $J_2$ the band minima are centred at the $\bs{K}$ points $\bs{K}_{\pm} = (\pm 4\pi/3,0)$. A long wavelength expansion of Eq.~\eqref{eq:haldane} yields the continuum Hamiltonian
\begin{align}
\hat{H}(\bs{K}_{\pm} + \bs{p}) &= -3 J_2 \cos \phi \; \hat{\sigma}_0 + \frac{\sqrt{3}}{2} J_1  (\mp p_x \hat{\sigma}_x - p_y \hat{\sigma}_y ) \nonumber \\ 
&+ (M - J_2 \sin \phi [ \pm \sqrt{3} + p_x + \sqrt{3} p_y ]) \hat{\sigma}_z, 
\end{align}
Here $M$ and $\phi$ have qualitatively different effects on the Dirac mass $M \mp J_2 \sqrt{3} \sin \phi$. When $\phi = 0$ the two valleys have the same mass, meaning that there is a small quantum distance between them, hence the low energy Bloch functions form a single connected cluster. On the other hand, $\phi$ creates two valleys with opposite masses, and hence eigenvectors residing on opposite sublattices with a large quantum distance between them, corresponding to two clusters. 

\revision{Even though the correspondence between the Betti numbers and the more familiar band topological invariants such as the Chern number $C$ is not exact, the latter may be used to guide the design of Bloch function clusters and loops. For example, in trivial $C=0$ phases the Bloch functions of a band can be continuously deformed to a point, such that the quantum distance vanishes~\cite{topo_review}. Hence the Bloch functions are likely to form a single cluster at any energy resolution. On the other hand, such a deformation is impossible in non-trivial $C \neq 0$ phases and there always exists Bloch functions separated by a large quantum distance, which enables the formation of distinct clusters. In the case of the Haldane model, phases with non-trivial Chern number $C$ occur when the two valleys have opposite signs of the Dirac mass.}


\section{Application to decay of quantum emitters}
\label{sec:emitter}

In this Section we will show how our chosen metrics can be used to predict the behaviour of interesting physical observables. Namely, we will demonstrate that the shape and connectivity of the energy band minima affect the lifetime of localized quantum emitters coupled to the lattice and resonant with the band edge, by imposing constraints on the effective emitter-Bloch wave coupling strength.

We consider the coupling of $N_e$ localized quantum emitters to the Haldane lattice, described by the Fano-Anderson model ($\hbar = 1$)~\cite{Longhi2018,Longhi2019,giant_atom,giant_atom2}
\begin{align}
    \hat{H} &= \hat{H}_e + \hat{H}_L + \hat{H}_{\mathrm{int}}, \quad \hat{H}_e = \sum_{\alpha=1}^{N_e} (\omega_{\alpha} - i \gamma_{\alpha}) \hat{c}_{\alpha}^{\dagger} \hat{c}_{\alpha}, \\
    \hat{H}_L &=  \sum_{n=1}^{N_b} \int d\bs{k} \omega_n(\bs{k}) \hat{u}_n(\bs{k}) \hat{u}_n^{\dagger}(\bs{k}),\\ \hat{H}_{\mathrm{int}} &= \sum_{\alpha,n} \int d\bs{k} g_{\alpha,n}(\bs{k}) \hat{c}_{\alpha}^{\dagger} \hat{u}_n(\bs{k}) + \mathrm{h.c.},
\end{align}
where $\hat{c}_{\alpha}$ annihilates a bound state with energy $\omega_{\alpha}$ and non-radiative decay rate $\gamma_{\alpha}$, $\hat{u}_n(\bs{k})$ annihilates a Bloch wave in band $n$ with momentum $\bs{k}$, and $g_{\alpha,n}(\bs{k})$ is the emitter-Bloch wave coupling strength,
\be 
g_{\alpha,n}(\bs{k}) = J_e \sum_{\bs{r}} \braket{ u_n(\bs{k}) | g_{\alpha}(\bs{r}) } e^{i \bs{k}\cdot \bs{r}} \equiv J_e \braket{ u_n(\bs{k}) | g_{\alpha}(\bs{k})},
\ee 
which we have decomposed into an energy scale $J_e$ and a dimensionless part $|\braket{ u_n(\bs{k}) | g_{\alpha}(\bs{k})}| = \sqrt{1 - d_{\alpha,n}(\bs{k})}$, where $d_{\alpha,n}(\bs{k})$ is the quantum distance between the emitter and Bloch wave polarizations at momentum $\bs{k}$. For simplicity, we focus on the single excitation subspace, in which the emitter and Bloch wave probability amplitudes evolve according to
\begin{align}
    i \partial_t c_{\alpha}(t) &= (\omega_{\alpha}-i\gamma_{\alpha}) c_{\alpha} + \int d\bs{k} g_{\alpha,n}(\bs{k}) u_n(\bs{k},t), \\
        i \partial_t u_n(\bs{k},t) &= \omega_n(\bs{k}) u_n(\bs{k},t) + \sum_{\alpha} g_{\alpha,n}^*(\bs{k}) c_{\alpha}(t).
\end{align}
We assume that only the emitters are excited at $t=0$, i.e. $u_n(\bs{k},0) = 0$, and applying the weak coupling and Markov approximations (see e.g. Refs.~\cite{Longhi2018,Longhi2019}), we arrive at an effective evolution equation for the emitter amplitudes, \be 
i \partial_t c_{\alpha}(t) = \sum_{\beta=1}^{N_e} \mathcal{H}_{\alpha,\beta} c_{\beta}(t),
\ee 
where $\mathcal{H}_{\alpha,\beta} = (\omega_{\alpha}-i\gamma_{\alpha}) \delta_{\alpha,\beta} + \Delta_{\alpha,\beta}$,
\be 
\Delta_{\alpha,\beta} = \int d\bs{k} \frac{g_{\alpha,1}(\bs{k}) g_{\beta,1}^*(\bs{k})}{\omega_1(\bs{k}) - \omega_{\beta}+ i \gamma_{\beta}} \label{eq:coupling}
\ee 
describes the emitters' coupling and self-energy mediated by the lattice, and we have dropped the summation over the band index $n$, as only the band energetically-closest to the emitters will be relevant in the weak coupling regime. Note that under the weak coupling approximation $J_e$ only sets the scale of $\Delta_{\alpha,\beta}$ and is otherwise irrelevant.

When the emitter frequencies are close to the lower band edge the integral in Eq.~\eqref{eq:coupling} will be dominated by modes with energies within the emitter linewidth $|\omega_1(\bs{k}) - \omega_{\beta}| < \gamma_{\beta}$, such that $\Delta_{\alpha,\beta}$ becomes sensitive to the shape and connectivity of the eigenvalues and Bloch functions within this energy resolution. For example, when the emitter couplings $g_{\alpha,1} \approx 0$ at the high symmetry points of the Brillouin zone, the emitter-lattice coupling will be strongly suppressed unless the band minimum forms a ``moat'' band. 

\begin{figure}
    \centering
    \includegraphics{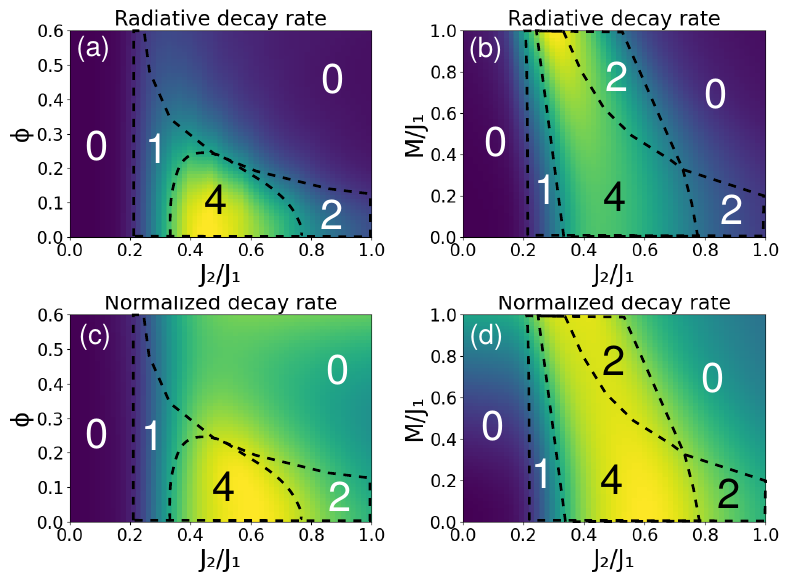}
    \caption{(a,b) Radiative decay rate $\Gamma$ of a ``giant atom'' coupled to a hexagonal plaquette of the Haldane model, described by the coupling Eq.~\eqref{eq:gfunction}. (c,d) Radiative decay rate normalized by the number of resonant modes Eq.~\eqref{eq:normalization}. Dashed lines and numbers denote the $B_1$ phase boundaries of Fig.~\ref{fig:haldane_figure_1}(c,d).}
    \label{fig:emitter_examples}
\end{figure}

We illustrate this in Fig.~\ref{fig:emitter_examples} for the case of a single emitter tuned to the band minimum with non-radiative decay rate $\gamma_e = 0.1J_1$ and coupled to a 6-site plaquette to form a ``giant atom''~\cite{giant_atom,giant_atom2}, described by the coupling
\be 
\ket{g_{e,e}(\bs{k})} = \left( \begin{array}{c} 1 + e^{-i \bs{k}\cdot\bs{a}_1} + e^{i \bs{k}\cdot \bs{a}_3} \\ e^{i \bs{k}\cdot\bs{\delta}_1} [1 + e^{-i \bs{k}\cdot \bs{a}_1} + e^{-i \bs{k}\cdot \bs{a}_2} ]\end{array} \right). \label{eq:gfunction}
\ee 
Fig.~\ref{fig:emitter_examples}(a,b) shows the radiative decay rate $\Gamma = -\mathrm{Im}[\Delta_{e,e}]$ as a function of $J_2,M$, and $\phi$. The radiative decay rate is smallest when $J_2 = 0$, corresponding to a point-like band minumum at $\bs{k}=0$. The peaks in the decay rate coincide reasonably well with the formation of the moat bands. To show that this enhancement of $\Gamma$ is sensitive to the shape of the low energy modes and not merely determined by the number of resonant modes, we also show in Fig.~\ref{fig:emitter_examples}(c,d) the normalized decay rate $\Gamma / \mathcal{N}$, where 
\be 
\mathcal{N} = J_e^2 \int d\bs{k} \left| \frac{\omega_1(\bs{k}) -\omega_{1} - i \gamma_{\beta}}{(\omega_1(\bs{k}) - \omega_{1})^2 + \gamma_{1}^2} \right|.\label{eq:normalization}
\ee
is the the $\ket{g_1(\bs{k})}$-independent part of Eq.~\eqref{eq:coupling}. The normalized decay rate is largest when there are multiple loops, i.e. $B_1 = 2$ or $4$, which optimizes the trade-off between emitter-Bloch wave mode detunings $\omega_1(\bs{k}) - \omega_e$ and the emitter-Bloch wave coupling $\braket{u_n(\bs{k})|g_e(\bs{k})}$.

\section{Conclusion}
\label{sec:conclusion}

We have shown how a method from the field of topological data analysis, namely persistent homology, can be used to identify novel classes of periodic lattices by systematically classifying the abstract shapes of their energy dispersion and Bloch functions. By using physically-motivated choices for the minimal persistence, we were able to eliminate discretization-induced noise in the topological features, thereby enabling reliable high-throughput classification of photonic band structures. \revision{The approach may be used to save many graduate student-hours, for example if it is necessary to compute and check thousands of band structures to find an optimal lattice design subject to certain constraints.} Moreoever, the method is highly flexible and can be readily generalized to higher dimensional settings~\cite{synthetic1,synthetic2}, complex lattices with large unit cells or overlapping bands such as Moir\'{e} superlattices~\cite{TBG, Moire_photonic}, continuous media such as photonic crystals~\cite{photonic_crystal}, and nonlinear or strongly interacting topological wave media~\cite{topo_review2}.

For the simple tight binding models we considered, the small parameter space could be covered through brute-force scanning. Persistent homology is compatible with gradient-based optimisation approaches, so an interesting direction for future research will be to use gradient-based methods to find specific energy band topologies of interest given constraints in the lattice design.

In contrast to some other recent studies employing persistent homology in physics~\cite{Murugan2019,Mengoni2019,Spitz2020,Tran2020,Olsthoorn2020,Cole_arxiv}, rather than focusing on the construction of persistence diagrams and their similarity measures, we focused on distance metrics directly related to physical observables of interest, such as the energy resolution. This allowed us to reliably separate robust features of the band structure from topological noise induced by numerical discretization. Future work may optimize the behaviour of other types of observables using persistent homology, \revision{study connections between the Betti numbers and the usual invariants of topological band theory such as the Chern number, and explore applications of persistent homology to other areas of photonics such as imaging.} We also hope this approach will also attract interest in branches of physics beyond photonics, such as the design and classification of novel condensed matter and quantum systems.

\section*{Acknowledgements}

This research was supported by the National Research Foundation, Prime Ministers Office, Singapore, the Ministry of Education, Singapore under the Research Centres of Excellence programme, and the Polisimulator project co-financed by Greece and the EU Regional Development Fund.

\section*{Data Availability}

The data that support the findings of this study are available from the corresponding author upon reasonable request.


\begin{thebibliography}{99}

\bibitem{Haldane2008}
\revision{F. D. M. Haldane and S. Raghu, {\it Possible Realization of Directional Optical Waveguides in Photonic Crystals with Broken Time-Reversal Symmetry}, Phys. Rev. Lett. {\bf 100}, 013904 (2008).}

 \bibitem{Wang2009}
 \revision{Z.  Wang,  Y.  Chong,  J.  D.  Joannopoulos,  and M. Solja\u{c}i\'{c}, {\it Observation of unidirectional backscattering-immune topological electromagnetic states}, Nature {\bf 461}, 772 (2009).}
 
 \bibitem{Malkova2009}
 \revision{N. Malkova, I. Hromada, X. Wang, G. Bryant, and Z. Chen, {\it Observation of optical Shockley-like surface states in photonic superlattices}, Opt. Lett. {\bf 34}, 1633 (2009).} 

\bibitem{Khanikaev2013}
\revision{A. B. Khanikaev, S. H. Mousavi, W.-K. Tse, M. Kargarian, A. H. MacDonald, and G. Shvets, {\it Photonic topological insulators}, Nature Materials {\bf 12}, 223 (2013).}

\bibitem{Rechtsman2013}
\revision{M. C. Rechtsman, J. M. Zeuner, Y. Plotnik, Y. Lumer, D. Podolsky, F. Dreisow, S. Nolte, M. Segev, and A. Szameit, {\it Photonic Floquet topological insulators}, Nature {\bf 496}, 196 (2013).}

\bibitem{Hafezi2013}
\revision{M. Hafezi, S. Mittal, J. Fan, A. Migdall, and J. M. Taylor, {\it Imaging topological edge states in silicon photonics}, Nature Photon. {\bf 7}, 1001 (2013).}

\bibitem{topo_review}
T. Ozawa, H. M. Price, A. Amo, N. Goldman, M. Hafezi, L. Lu, M. C. Rechtsman, D. Schuster, J. Simon, O. Zilberberg, and I. Carusotto, {\it Topological photonics}, Rev. Mod. Phys. {\bf 91}, 015006 (2019).

\bibitem{topo_review2}
D. Smirnova, D. Leykam, Y. Chong, and Y Kivshar, {\it Nonlinear topological photonics}, Appl. Phys. Rev. {\bf 7}, 021306 (2020).

\bibitem{TI_review}
M. Z. Hasan and C. L. Kane, {\it Colloquium: Topological insulators}, Rev. Mod. Phys. {\bf 82}, 3045 (2010).

\bibitem{synthetic1}
T. Ozawa, H. M. Price, N. Goldman, O. Zilberberg, and I. Carusotto, {\it Synthetic dimensions in integrated photonics: From optical isolation to four-dimensional quantum Hall physics}, Phys. Rev. A {\bf 93}, 043827 (2016).

\bibitem{synthetic2}
L. Yuan, Q. Lin, M. Xiao, and S. Fan, {\it Synthetic dimension in photonics}, Optica {\bf 5},  1396 (2018).

\bibitem{synthetic3}
L. J. Maczewsky, K. Wang, A. A. Dovgiy, A. E. Miroshnichenko, A. Moroz, M. Ehrhardt, M. Heinrich, D. N. Christodoulides, A. Szameit, and A. A. Sukhorukov, {\it Synthesizing multi-dimensional excitation dynamics and localization transition in one-dimensional lattices}, Nature Photonics {\bf 14}, 76 (2020).

\bibitem{hyperbolic_review}
A. Poddubny, I. Iorsh, P. Belov, and Y. Kivshar, {\it Hyperbolic metamaterials}, Nature Photon. {\bf 7}, 948 (2013). 

\bibitem{FB_review}
D. Leykam and S. Flach, {\it Perspective: Photonic flatbands}, APL Photonics {\bf 3}, 070901 (2018).

\bibitem{Ghrist2008}
R. Ghrist, {\it Barcodes: The persistent topology of data}, Bull. Amer. Math. Soc. {\bf 45}, 61 (2008).

\bibitem{TDA_review}
L. Wasserman, {\it Topological Data Analysis}, Annual Review of Statistics and Its Application, {\bf 5}, 501 (2017).

\bibitem{TDA_review2}
G. Carlsson, {\it Topological methods for data modelling}, Nature Rev. Phys. {\bf 2}, 697 (2020).

\bibitem{Murugan2019}
J. Murugan, D. Robertson, {\it An Introduction to Topological Data Analysis for Physicists: From LGM to FRBs}, arXiv:1904.11044

\bibitem{Mengoni2019}
R. Mengoni, A. Di Pierro, L. Memarzadeh, S. Mancini, {\it Persistent homology analysis of multiqubit entanglement}, Quantum Information and Computation {\bf 20}, 0375 (2020).

\bibitem{Spitz2020}
D. Spitz, J. Berges, M. K. Oberthaler, A. Wienhard, {\it Finding universal structures in quantum many-body dynamics via persistent homology}, arXiv:2001.02616.

\bibitem{Tran2020}
Q. H. Tran, M. Chen, and Y. Hasegawa, {\it Topological Persistence Machine of Phase Transitions}, arXiv:2004.03169.

\bibitem{Olsthoorn2020}
B. Olsthoorn, J. Hellsvik, and A. V. Balatsky, {\it Finding hidden order in spin models with persistent homology}, Phys. Rev. Res. {\bf 2}, 043308 (2020).

\bibitem{Cole_arxiv}
A. Cole, G. J. Loges, and G. Shiu, {\it Quantitative and Interpretable Order Parameters for Phase Transitions from Persistent Homology}, arXiv:2009.14231.

\bibitem{Haldane0}
F. D. M. Haldane, {\it Model for a Quantum Hall Effect without Landau Levels: Condensed-Matter Realization of the ``Parity Anomaly''}, Phys. Rev. Lett. {\bf 61}, 2015 (1988).

\bibitem{Haldane1}
M. Minkov and V. Savona, {\it Haldane quantum Hall effect for light in a dynamically modulated array of resonators}, Optica {\bf 3}, 200 (2016).

\bibitem{Haldane2}
S. Mittal, V. V. Orre, D. Leykam, Y. D. Chong, and M. Hafezi {\it Photonic Anomalous Quantum Hall Effect}, Phys. Rev. Lett. {\bf 123}, 043201 (2019).

\bibitem{Haldane3}
S. Lanneb\`{e}re and M. G. Silveirinha, {\it Photonic analogues of the Haldane and Kane-Mele models}, Nanophotonics {\bf 8}, 1387 (2019).

\bibitem{Haldane_talk}
F. D. M. Haldane, {\it Dirac-point models: Hilbert space geometry and topology} \url{http://wwwphy.princeton.edu/~haldane/talks/nobel_jpeg.pdf} (2010).

\bibitem{Kolodrubetz2013}
M. Kolodrubetz, V. Gritsev, and A. Polkovnikov, {\it Classifying and measuring geometry of a quantum ground state manifold}, Phys. Rev. B {\bf 88}, 064304 (2013).

\bibitem{Palumbo2018}
G. Palumbo, {\it Momentum-space cigar geometry in topological phases}, Eur. Phys. J. Plus {\bf 133}, 23, (2018).

\bibitem{Bleu2018}
O. Bleu, G. Malpuech, Y. Gao, and D. D. Solnyshkov, {\it Effective Theory of Nonadiabatic Quantum Evolution Based on the Quantum Geometric Tensor}, Phys. Rev. Lett. {\bf 121}, 020401 (2018).

\bibitem{Rhim2020}
J.-W. Rhim, K. Kim, and B.-J. Yang, {\it Quantum distance and anomalous Landau levels of flat bands}, Nature {\bf 584}, 59 (2020).

\bibitem{Gianfrate2020}
A. Gianfrate, O. Bleu, L. Dominici, V. Ardizzone, M. De Giorgi, D. Ballarini, G. Lerario, K. W. West, L. N. Pfeiffer, D. D. Solnyshkov, D. Sanvitto, and G. Malpuech, {\it Measurement of the quantum geometric tensor and of the anomalous Hall drift}, Nature {\bf 578}, 381 (2020).

\bibitem{SCQ}
X. Tan et al., {\it Experimental Measurement of the Quantum Metric Tensor and Related Topological Phase Transition with a Superconducting Qubit}, Phys. Rev. Lett. {\bf 122}, 210401 (2019).

\bibitem{Sedrakyan2014}
T. A. Sedrakyan, L. I. Glazman, and A. Kamenev, {\it Absence of Bose condensation on lattices with moat bands}, Phys. Rev. B {\bf 89}, 201112(R) (2014).

\bibitem{Sedrakyan2015}
T. A. Sedrakyan, L. I. Glazman, and A. Kamenev, {\it Spontaneous Formation of a Nonuniform Chiral Spin Liquid in a Moat-Band Lattice}, Phys. Rev. Lett. {\bf 114}, 037203 (2015).

\bibitem{multivalley}
M. Sun, I. G. Savenko, H. Flayac, and T. C. H. Liew, {\it Multivalley engineering in semiconductor microcavities}, Scientific Reports {\bf 7}, 45243 (2017).

\bibitem{multivalley2}
H. S. Nguyen, F. Dubois, T. Deschamps, S. Cueff, A. Pardon, J.-L. Leclercq, C. Seassal, X. Letartre, and P. Viktorovitch, {\it Symmetry Breaking in Photonic Crystals: On-Demand Dispersion from Flatband to Dirac Cones}, Phys. Rev. Lett. {\bf 120}, 066102 (2018).


\bibitem{Julia}
J. Bezanson, A. Edelman, S. Karpinski, V. B. Shah, {\it Julia: A Fresh Approach to Numerical Computing}, SIAM Review {\bf 59}, 65 (2017).

\bibitem{Eirene}
G. Henselman, R. Ghrist, {\it Matroid Filtrations and Computational Persistent Homology}, arXiv:1606.00199.

\bibitem{Ripserer}
U. Bauer, {\it Ripser: efficient computation of Vietoris-Rips persistence barcodes}, arXiv:1908.02518.

\bibitem{Longhi2018}
S. Longhi, {\it Anomalous dynamics in multilevel quantum decay}, Phys. Rev. A {\bf 98}, 022134 (2018).

\bibitem{Longhi2019}
S. Longhi, {\it Quantum decay in a topological continuum}, Phys. Rev. A {\bf 100}, 022123 (2019).

\bibitem{giant_atom}
A. F. Kockum, G. Johansson, and F. Nori, {\it Decoherence-Free Interaction between Giant Atoms in Waveguide Quantum Electrodynamics}, Phys. Rev. Lett. {\bf 120}, 140404 (2018).

\bibitem{giant_atom2}
A. F. Kockum, {\it Quantum optics with giant atoms -- the first five years}, in International Symposium on Mathematics, Quantum Theory, and Cryptography, Springer, Singapore (2020).

\bibitem{TBG}
R. Bistritzer and A. H. MacDonald, {\it Moire bands in twisted double-layer graphene}, PNAS {\bf 108}, 12233 (2011).

\bibitem{Moire_photonic}
P. Wang, Y. Zheng, X. Chen, C. Huang, Y. V. Kartashov, L. Torner, V. V. Konotop, and F. Ye, {\it Localization and delocalization of light in photonic Moire lattices}, Nature {\bf 577}, 42 (2020).

\bibitem{photonic_crystal}
J. D. Joannopoulos, S. G. Johnson, J. N. Winn, and R. D. Meade, {\it Photonic Crystals: Molding the Flow of Light}, Princeton University Press (2008).


\end{thebibliography}
\end{document}